\documentclass[letterpaper, 10 pt, conference]{ieeeconf}  

\IEEEoverridecommandlockouts                              

\usepackage{graphics} 
\usepackage{graphicx}
\usepackage{epsfig} 
\usepackage{mathptmx} 
\usepackage{times} 
\usepackage{amsmath} 
\usepackage[T1]{fontenc}
\usepackage{amsfonts,amssymb}
\usepackage{bm}

\usepackage{CJK}
\usepackage{algpseudocode}
\newcommand{\RNum}[1]{\uppercase\expandafter{\romannumeral #1\relax}}

\title{\LARGE \bf
Safe Reinforcement Learning for a Robot Being Pursued but with Objectives Covering More Than Capture-avoidance

}

\author{Huanhui Cao$^{1}$, Zhiyuan Cai$^{1}$, Hairuo Wei$^{1}$, Wenjie Lu$^{1}$, Lin Zhang$^{2}$, and Hao Xiong$^{1,\dag}$

\thanks{$^{1}$Huanhui Cao, Zhiyuan Cai, Hairuo Wei, Wenjie Lu, and Hao Xiong are with the School of Mechanical Engineering and Automation, Harbin Institute of Technology Shenzhen, Shenzhen, China.
        }%
\thanks{$^{2}$Lin Zhang is with the Department of Aerospace Engineering and Engineering Mechanics, University of Cincinnati, Cincinnati, OH 45040, USA.
        }%
        
\thanks{$\dag$Corresponding author: e-mail: xionghao@hit.edu.cn.}
}

\begin{document}

\maketitle
\thispagestyle{empty}
\pagestyle{empty}

\begin{abstract}
Reinforcement Learning (RL) algorithms show amazing performance in recent years, but placing RL in real-world applications such as self-driven vehicles may suffer safety problems. A self-driven vehicle moving to a target position following a learned policy may suffer a vehicle with unpredictable aggressive behaviors or even being pursued by a vehicle following a Nash strategy. To address the safety issue of the self-driven vehicle in this scenario, this paper conducts a preliminary study based on a system of robots. A safe RL framework with safety guarantees is developed for a robot being pursued but with objectives covering more than capture-avoidance.
Simulations and experiments are conducted based on the system of robots to evaluate the effectiveness of the developed safe RL framework.

\end{abstract}

\section{Introduction \label{intro}}

\begin{figure}[htb]
    \centering
    \includegraphics[width=4.5cm]{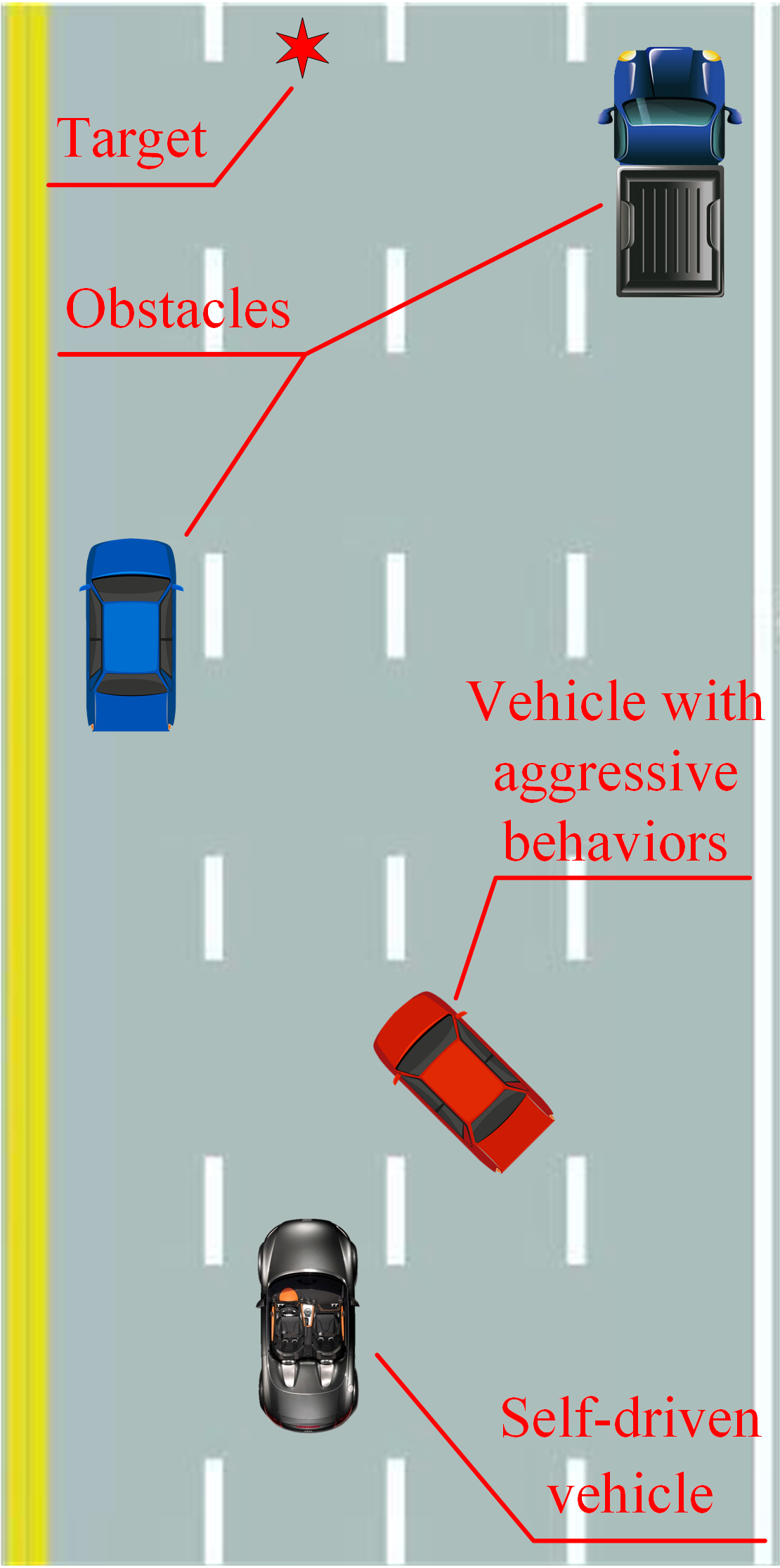}
    \caption{A self-driven vehicle suffering a vehicle with unpredictable aggressive behaviors}
    \label{fig:vehicles}
\end{figure}

Reinforcement learning (RL) has achieved dramatic success in many areas such as robot manipulation \cite{Lillicrap2015} and playing Go \cite{Silver2016} in recent years. Safety is critical to many applications of RL, such as self-driving vehicles and robot manipulation \cite{Ames2017ControlSystems,Berkenkamp2016SafeProcesses}. For safety-critical applications such as self-driven vehicles, the failure of a learned policy could result in collisions of vehicles and injury or death of humans \cite{Xiong2021SafetyControl}. 
In practice, a self-driven vehicle can suffer vehicles with unpredictable aggressive behaviors, as shown in Fig. \ref{fig:vehicles}. To ensure the self-driven vehicle is free of a collision with a vehicle with unpredictable aggressive behaviors, a learned policy for the self-driven vehicle should be effective even in the worst-case, if it is possible. Namely, even if the vehicle with unpredictable aggressive behaviors actually pursues the self-driven vehicle following a Nash strategy \cite{Lin2015NashObservations}, a learned policy for the self-driven vehicle should avoid capture. However, except for capture-avoidance, a learned policy for the self-driven vehicle usually includes objects other than capture-avoidance (e.g., reaching a target position). Thus, a Nash strategy for the evader of a pursuit-evasion game cannot be applied to a learned policy for the self-driven vehicle to achieve objectives covering more than capture-avoidance.
\begin{figure}[htb]
    \centering
    \includegraphics[width=7.6cm]{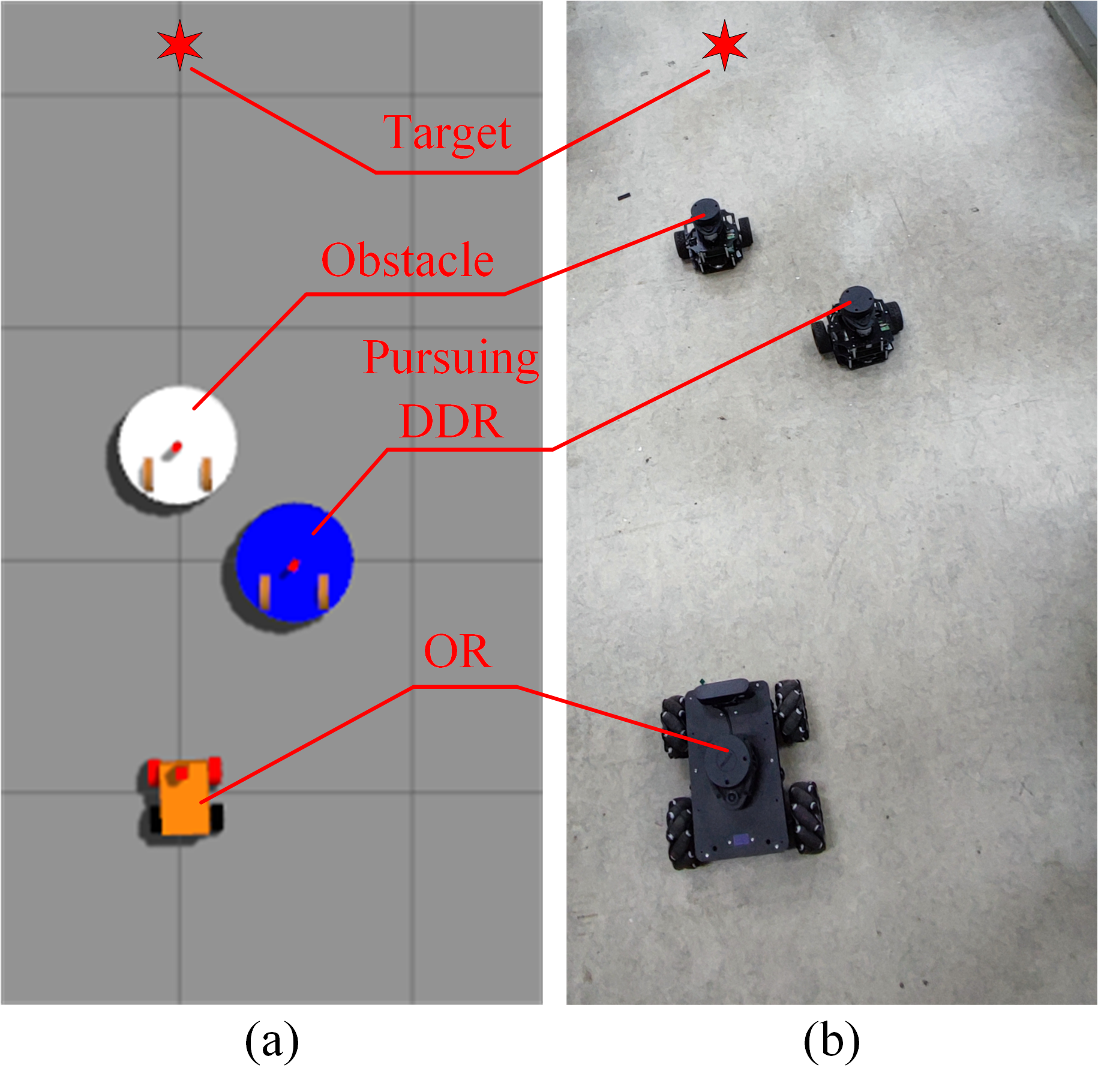}
    \caption{An OR suffering a DDR with unpredictable aggressive behaviors a) in simulation and b) in real-world}
    \label{fig:experiment}
\end{figure}

To address the safety issue of a learned policy for a self-driven vehicle suffering vehicles with unpredictable aggressive behaviors, this paper proposes a problem for robots - \emph{How a robot being pursued can guarantee capture-avoidance and achieve objectives covering more than capture-avoidance based on RL?} To solve this problem, a preliminary study is conducted based on a system of robots, as shown in Fig. \ref{fig:experiment}. An Omnidirectional Robot (OR) plays the role of a self-driven vehicle. A learned policy for the OR needs to 1) reach a randomly generated target position based RL; 2) avoid being captured, assuming that capture-avoidance is possible; and 3) avoid obstacles. A Differential Drive Robot (DDR) plays the role of a vehicle with unpredictable aggressive behaviors. The DDR pursues the OR following a Nash strategy. Another DDR plays the role of a moving obstacle.

Achievements in pursuit-evasion games are valuable for this study. The pursuit-evasion game of capturing an omnidirectional evader using a differential drive robot was studied in \cite{Ruiz2013Time-OptimalRobot}. Time-optimal Nash strategies were obtained for both the pursuer and the evader. It was shown that any unilateral deviation of the pursuer or the evader from Nash strategies does not provide a benefit to win the game. Macias et. al analyzed the value of velocity information on the pursuit-evasion game of capturing an omnidirectional evader using a differential drive robot in \cite{Macias2018ImageGame}. Base on \cite{Ruiz2013Time-OptimalRobot,Macias2018ImageGame}, one can determine the worst-case and the possibility of capture-avoidance for an OR being perused by a DDR.

Safe RL takes the safety issue into account, aiming to learn a policy that maximizes the expected reward on the condition that safety constraints are satisfied \cite{Garcia2012SafeLearning}. Several safe RL approaches have been proposed, including learning from demonstration, policy optimization with constraints, and reward-shaping. However, the major issue of these approaches is that safety is not guaranteed during initial learning interactions \cite{Cheng2019End-to-endTasks}. To address this issue, shielding frameworks have been proposed, including shield RL \cite{Alshiekh2018SafeShielding}. According to the shielding frameworks, shields are synthesized to guarantee safety during learning by monitoring the actions of agents. However, it can be challenging to design shields in certain applications.

It is proposed in \cite{Recht2019AControl} that solving problems in safety-critical applications requires the deep fusion of both machine learning and control technologies. Control Barrier Function (CBF) based methods can lead to effective shields in safety-critical applications \cite{Ames2019ControlApplications}. Safe RL approaches based on integrating RL with shields constructed by CBFs have been proposed in \cite{Cheng2019End-to-endTasks,Marvi2021SafeApproach}. To ensure safety in multi-agent problems, multi-agent CBF has been developed in \cite{Borrmann2015ControlBehavior}. Different CBFs have been designed assuming that agents are cooperative, neutral, and competitive, respectively. Cheng et al. \cite{Cheng2020SafeUncertainties} have further improved the multi-agent CBF from the perspective of uncertainty bounds. However, CBFs have been designed for pursuit-evasion games in previous studies. 

In view of the safety issue of a learned policy for a self-driven vehicle suffering vehicles with unpredictable aggressive behaviors, this paper summaries a safe RL problem for a robot being pursued. To address the safe RL problem, this paper develops a safe RL framework by integrating RL with shields based on CBFs. The major contributions of this paper are as follows.

\begin{itemize}
\item This paper proposes a safe RL problem for robots being perused but with objectives covering more than capture-avoidance. By integrating RL with shields constructed by CBFs, a safe RL framework with safety guarantees is developed for robots to address the proposed safe RL problem.

\item A CBF is designed to provide capture-avoidance guarantees for an OR being pursued by a DDR following a Nash strategy, if it is possible for the OR to avoid being captured.

\item Simulations and experiments of a system of robots, consisting of an OR, a pursuing DDR, and an obstacle, are conducted to evaluate the effectiveness of the developed safe RL framework.

\end{itemize}

The rest of the paper is organized as follows. Section \ref{pre} introduces the preliminaries of this paper. In Section \ref{td3_cbf}, a safe RL framework with safety guarantees is developed for a robot being pursued but with objectives covering more than capture-avoidance. The design of CBFs for an OR suffering a pursuing DDR and obstacles is presented. In Section \ref{experiment}, the effectiveness of the developed safe RL framework is evaluated based on simulations andexperiments. Finally, Section \ref{summary} summarizes this paper.

\section{Preliminaries \label{pre}}
In this section, the system of robots discussed in this paper, reinforcement learning, and control barrier function are introduced briefly.

\subsection{A System of Robots \label{pro_formu}}

In this paper, the safe RL of a robot being pursued but with objectives covering more than capture-avoidance is discussed based on a system of robots including an OR, a DDR, and an obstacle, as shown in Fig. \ref{fig:experiment}. The OR needs to reach a random target position based on a learned policy; 2) avoid being captured; and 3) avoid obstacles. The DDR pursues the OR following a Nash strategy obtained according to \cite{Ruiz2013Time-OptimalRobot}. The obstacle moves at a constant velocity.

The robots can be modeled in the Euclidean plane according to \cite{Ruiz2013Time-OptimalRobot}. 
Let $(x_d, y_d, \theta_d)$ presents the pose of the DDR. $(x_o, y_o)$ and $(x_c, y_c)$ denote the position of the OR and the position of an obstacle with a constant velocity, respectively. 
The motion equation of the OR can be formulated as
\begin{equation}
    \label{dynamics_OA}
    \begin{cases}
    \dot{x}_o = v_o \cos{\theta_o}\\
     \dot{y}_o = v_o \sin{\theta_o}\\
    \end{cases}
\end{equation}
It is supposed that the translational velocity $v_o$ is a constant and the steering angle $\theta_o$ is directly specified according to \cite{Ruiz2013Time-OptimalRobot}.
The motion equation of the DDR can be formulated as
\begin{equation}
    \label{dynamics}
    \begin{cases}
    \dot{x}_d = (\frac{u_1+u_2}{2})\cos{\theta_d} \\
    \dot{y}_d =  (\frac{u_1+u_2}{2})\sin{\theta_d}\\
    \dot{\theta_d} = (\frac{u_2-u_1}{2b})\\
    \end{cases}
\end{equation}
where $\theta_d$ is the heading angular of the DDR. $u_1$ and $u_2$ are the wheel angular velocities of the DDR. $b$ is the distance between the center of the DDR and the wheel location. The translational velocity of the DDR, denoted as $v_d$, satisfies
\begin{equation}
    v_d=\frac{u_1+u_2}{2}
\end{equation}
The motion equation of the obstacle with a constant velocity can be formulated as
\begin{equation}
    \label{dynamics_obstacle}
    \begin{cases}
    \dot{x}_c = v_{cx} \\
     \dot{y}_c = v_{cy} \\
    \end{cases}
\end{equation}
where $v_{cx}$ and $v_{cy}$ and are the translational velocity component of the obstacle along the $x$ axes and that of the obstacle along the $y$ axes, respectively.
It should be noted that $v_d$, $v_o$, $v_{cx}$, and $v_{cy}$ are constants in this study.


\subsection{Reinforcement Learning}
Reinforcement learning is an approach for an agent to achieve a learned policy by maximizing the expected cumulative rewards when interacting with the environment \cite{Henderson2018}. At time step $t$, the agent selects an action $u \in U$ based on the current state $x \in X$ with respect to its policy $\pi: X \mapsto U$. The agent receives a reward $r$ and the state $x$ transfers to a new state $x'$. The agent aims to maximize the accumulated rewards $R_t = \sum_{i=t}^{t_{max}}{\gamma^{i-t}r(x_i,u_i)}$, where $\gamma$ is a discount factor.

\subsection{Control Barrier Function}
A control barrier function plays a role in the study of safety equivalent to a Lyapunov function in the study of stability \cite{Ames2019ControlApplications}. Without loss of generality, one can assume a nonlinear affine system
\begin{equation}
\label{controlSystem}
\dot{x} = f(x) + g(x)u
\end{equation}with $f$ and $g$ locally Lipschitz, $x \in X \subset \mathbb{R}^n $ and $u \in U \subset \mathbb{R}^m$. Safety of the system can be guaranteed via enforcing the invariance of a safe set \cite{Ames2019ControlApplications}. In particular, one can consider a set $C$ defined as the superlevel of a continuously differentiable function $h(x):\mathbb{R}^n \rightarrow \mathbb{R}$, yielding
\begin{equation}
\begin{split}
\label{setC}
C=\{x\in \mathbb{R}^n : h(x)\ge 0\}\\
\partial C=\{x\in  \mathbb{R}^n : h(x) = 0\}\\
Int(C)=\{x\in \mathbb{R}^n : h(x) > 0\}
\end{split}
\end{equation}
We refer to $C$ as the \emph{\underline{safe set}}.

\noindent \textbf{Definition 1.} \cite{Ames2019ControlApplications} \emph{The set $C$ is \underline{forward invariant} if for every $x(0) \in C, x(t) \in C $ for all $t \in [0, t_{max})$. The system (\ref{controlSystem}) is \underline{safe} with respect to the set $C$ if the set $C$ is forward invariant.}

\noindent \textbf{Definition 2.} \cite{Ames2019ControlApplications} \emph{For the system (\ref{controlSystem}), a continuously differentiable function $h(x):\mathbb{R}^n\rightarrow\mathbb{R}$ is a \underline{control barrier function} for the set $C$ defined by (\ref{setC}) if there exist locally Lipschitz class $\kappa_{\inf}$ function $\alpha$ such that, for all $x\in Int(C)$},
\begin{equation}
    \sup_{u \in U}[L_fh(x)+L_gh(x)u] \ge -\alpha(h(x))
    \label{hx}
\end{equation}
Given a CBF $h(x)$ and a set of Lipschitz continuous controller
\begin{equation}
    K_{cbf}(x) = \{u\in U:L_fh(x)+L_gh(x)u +\alpha(h(x)) \ge 0 \}
\end{equation}
one has

\noindent \textbf{Theorem 1.} \cite{Ames2019ControlApplications} \emph{Given a set $C\subset \mathbb{R}^n$ defined by (\ref{setC}) with associated control barrier function h(x), any Lipschitz continuous controller $u \in K_{cbf}(x)$ for the system (\ref{controlSystem}) renders the set $C$ \underline{forward invariant}, ensuring that the system (\ref{controlSystem}) is  \underline{safe} with respect to the set $C$.}

\section{Safe RL for an OR Being Pursued by a DDR but with Objectives Covering More Than Capture-avoidance \label{td3_cbf}}
To address the safe RL problems of a robot being pursued but with objectives covering more than capture-avoidance, a safe RL framework is established for a robot in this section. To apply the proposed safe RL framework to the OR included in the system of robots discussed in this paper, CBFs are designed for the OR to avoid being captured and to avoid obstacles. 

\subsection{Safe RL Framework for a Robot Being Pursued but with Objectives Covering More Than Capture-avoidance}
\begin{figure}[h]
    \centering
    \includegraphics[width=8cm]{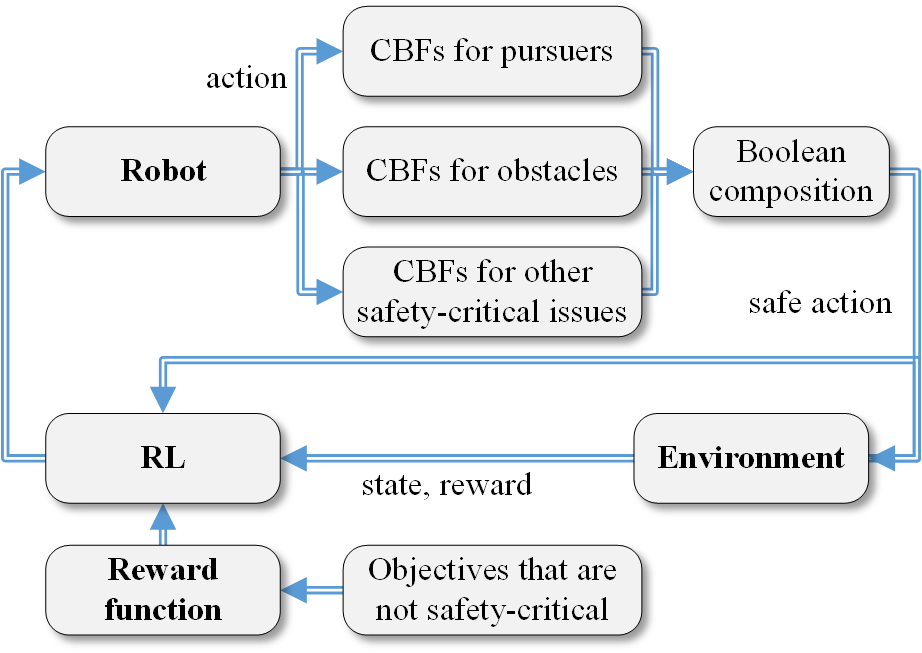}
    \caption{A safe reinforcement learning framework for a robot being pursued but with objectives covering more than capture-avoidance}
    \label{fig:framework}
\end{figure}
It is proposed in \cite{Cheng2019End-to-endTasks,Marvi2021SafeApproach} to integrate RL with shields based on CBFs to address the safety issues of a general RL that does not have a specific mechanism for ensuring safety. In this paper, a safe RL framework is established for a robot being pursued but with objectives covering more than capture-avoidance by integrating RL with multiple shields based on CBFs, as shown in Fig. \ref{fig:framework}. According to Fig. \ref{fig:framework}, capture-avoidance, obstacle-avoidance, and other safety-critical issues are addressed by CBFs. The action of a robot is corrected by CBFs, if necessary, and a safe action is performed. Objectives that are not safety-critically can be integrated into the reward function of RL. 

For a robot has $k$ CBFs (i.e., $h_i(x) i=1,2,3,...,k$), the CBFs can be combined through Boolean composition according to \cite{Glotfelter2017NonsmoothSystems} and achieves a composite CBF $h_c(x)$ as
\begin{equation}
\label{Boolean}
\begin{split}
    h_c(x) &= h_1(x) \cap  h_2(x)\cap ... \cap  h_k(x) \\
\end{split}
\end{equation}
It is expected that the composite CBF lead the minimal interference \cite{Fisac2019ASystems,ElSayed-Aly2021SafeShielding} to the action of a robot. Namely, 1) the composite CBF corrects the action of a robot only if it tends to violate safety constraints, and 2) the composite CBF revises the action of a robot as few as possible. A composite CBF derived based on (\ref{Boolean}) can lead to a safe action $u_{safe}$ for a robot as \cite{Ames2019ControlApplications}
\begin{equation}
\label{qp}
\begin{split}
&u_{safe}= \arg\min_u \frac{1}{2}\vert\vert u-\hat u\vert\vert^2\\
s.t. 
 \ \ \ & L_fh(x)+L_gh(x)u +\alpha(h(x)) \ge 0 \\
 \ \ \  & \quad \quad \quad\quad\vert\vert u \vert \vert\leq u_{max} 
\end{split}
\end{equation}
where $\hat{u}$ is the nominal action of the robot following a learned policy. $u_{max}$ is the limit of the action of the robot.



\subsection{Control Barrier Function for an Obstacle \label{cbf_eo}}
To apply the proposed safe RL framework to the OR of the system of robots discussed in this paper, it is significant to design the CBFs of the OR. In this paper, the CBFs of the OR for obstacles and the CBF of the OR for a pursuing DDR are designed. 

Based on the assumption that the position and the constant velocity of the $i$th obstacle are available for the OR, the CBF of the OR for the $i$th obstacle is designed. The position of the $i$th obstacle is denoted as $(x_{c}^i,y_{c}^i)$. $v^i_{cx}$ and $v^i_{cy}$ and are the constant translational velocity component of the $i$th obstacle along the $x$ axes and that of the $i$th obstacle along the $y$ axes, respectively. 
The collision distance between the OR and an obstacle is $d_{oc}$. 
Let $\Delta p^i_{oc}$  represents the difference between the position of the OR and that of the $i$th obstacle. $\Delta v^i_{oc}$ represents the difference between the velocity of the OR and that of the $i$th obstacle. One has
\begin{equation}
\begin{cases}
 &\Delta p_{oc}^i = [x_o-x_{c}^i,y_o-y_{c}^i]^T\\
    &\Delta v_{oc}^i = [v_o \cos{\theta_o}-v^i_{cx}, v_o \sin{\theta_o}-v^i_{cy}]^T
    \end{cases}
\end{equation}
For the $i$th obstacle, a safe set is defined as
\begin{equation}
    C_{c}^i=\left\{(x_o,y_o) \ | \ \lvert \lvert \Delta p_{oc}^i \rvert \rvert-d_{oc}\ge 0\right \}
\end{equation}
A continuously differentiable function $h_{oc}^i(x)$ is defined as
\begin{equation}
\label{h_obstacle}
    h_{oc}^i(x)=\lvert \lvert \Delta p_{oc}^i \rvert \rvert-d_{oc}
\end{equation}
According to (\ref{hx}), let $\alpha(x) = \gamma^i_{oc} x$. $\gamma^i_{oc}$ is a  tunable parameter and $\gamma^i_{oc} >0$. Then one has
\begin{equation}
\label{Au1}
\begin{split}
&L_fh_{oc}^i(x)+L_gh_{oc}^i(x)u^i_{oc} +\alpha(h_{oc}^i(x))\\
&=\frac{\Delta p_{oc}^{i,T} \Delta v^i_{oc}}{\lvert \lvert \Delta p_{oc}^{i} \rvert \rvert } +\gamma_{oc}^i h_{oc}^i(x)\\
&=\frac{\Delta p_{oc}^{i,T} }{\lvert \lvert \Delta p_{oc}^{i} \rvert \rvert } \Delta v^i_{oc}+\gamma^i_{oc}(\lvert \lvert \Delta p_{oc}^i \rvert \rvert-d_{oc})\\
&=\frac{\Delta p_{oc}^{i,T} }{\lvert \lvert \Delta p_{oc}^{i} \rvert \rvert } [v_o \cos{\theta_o}-v^i_{cx}, v_o \sin{\theta_o}-v^i_{cy}]^T+\gamma^i_{oc}(\lvert \lvert \Delta p_{oc}^i \rvert \rvert-d_{oc}) \\
&=\frac{\Delta p_{oc}^{i,T} }{\lvert \lvert \Delta p_{oc}^{i} \rvert \rvert } [v_o \cos{\theta_o}, v_o \sin{\theta_o}]^T- \frac{\Delta p_{oc}^{i,T} }{\lvert \lvert \Delta p_{oc}^{i} \rvert \rvert }[v^i_{cx}, v^i_{cy}]^T \\
&\quad+\gamma^i_{oc}(\lvert \lvert \Delta p_{oc}^i \rvert \rvert-d_{oc}) \\
&=\frac{\Delta p_{oc}^{i,T} }{\lvert \lvert \Delta p_{oc}^{i} \rvert \rvert } u_{oc}^i-\frac{\Delta p_{oc}^{i,T} }{\lvert \lvert \Delta p_{oc}^{i} \rvert \rvert }[v^i_{cx}, v^i_{cy}]^T+\gamma^i_{oc}(\lvert \lvert \Delta p_{oc}^i \rvert \rvert-d_{oc}) \ge 0\\
&=-A_{oc}^iu_{oc}^i+b_{oc}^i \ge 0
\end{split}
\end{equation}
where
\begin{equation}
\begin{cases}
A_{oc}^i=\frac{\Delta p_{oc}^{i,T} }{\lvert \lvert \Delta p_{oc}^{i} \rvert \rvert }\\
b_{oc}^i=-\frac{\Delta p_{oc}^{i,T} }{\lvert \lvert \Delta p_{oc}^{i} \rvert \rvert }[v^i_{cx}, v^i_{cy}]^T+\gamma^i_{oc}(\lvert \lvert \Delta p_{oc}^i \rvert \rvert-d_{oc})\\
u^i_{oc} = [v_o \cos{\theta_o}, v_o \sin{\theta_o}]^T
\end{cases}
\end{equation}
According to \textbf{Theorem 1}, if $u_{oc}^i$  satisfies (\ref{Au1}) and $\lvert\lvert u_{oc}^i\rvert\rvert \leq u_{oc}^{max}$, then $h_{oc}^i(x)$ is a valid CBF and the safe set $C_{oc}^i$ is forward invariant, ensuring  that the OR will not collide with the $i$th obstacle. With an appropriate selection of $\gamma_{oc}^i$, one can make $u_{oc}^i$ satisfies (\ref{Au1}).  

\subsection{Control Barrier Function for a Pursuing DDR \label{cbf_pv}}


Based on the assumption that the position and the heading angle of a pursuing DDR are available for an OR, the CBF of the OR for the pursuing DDR is designed.
Let $\Delta p_{pv}$ represents the difference between the position of the OR and the position of the DDR. $\Delta v_{pv}$ denotes the difference between the velocity of the OR and the velocity of the DDR. $d_{pv}$ represents the capture distance between the DDR and the OR. One has 
\begin{equation}
\begin{cases}
    \Delta p_{pv} = [x_o-x_d,y_o-y_d]^T\\
    \Delta v_{pv} = [v_o \cos{\theta_o}-v_d\cos{\theta_d}, v_o \sin{\theta_o}-v_d\sin{\theta_d}]^T
\end{cases}
\end{equation}
A safe set is defined as
\begin{equation}
      C_{pv}=\left\{(x_o,y_o) \ | \ \lvert \lvert \Delta p_{pv} \rvert \rvert-d_{pv}\ge 0\right \}
\end{equation}
where $d_{pv}$ is the safety distance. A continuously differentiable function $h_{pv}(x)$ is defined as
\begin{equation}
\label{h_pv}
      h_{pv}(x)=\lvert \lvert \Delta p_{pv} \rvert \rvert-d_{pv}
\end{equation}
According to (\ref{hx}), let $\alpha(x) = \gamma_{pv} x$ ($\gamma_{pv} >0$). Then one has
\begin{equation}
\label{Au2}
\begin{split}
&L_fh_{pv}(x)+L_gh_{pv}(x)u_{pv} +\alpha(h_{pv}(x))\\
&=\frac{\Delta p_{pv}^{T} \Delta v_{pv}}{\lvert \lvert \Delta p_{pv} \rvert \rvert } +\gamma_{pv}h_{pv}(x)\\
&=\frac{\Delta p_{pv}^{T} }{\lvert \lvert \Delta p_{pv} \rvert \rvert } \Delta v_{pv}+\gamma_{pv}(\lvert \lvert \Delta p_{pv} \rvert \rvert-d_{pv}) \\
&=\frac{\Delta p_{pv}^{T} }{\lvert \lvert \Delta p_{pv} \rvert \rvert } [v_o \cos{\theta_o}-v_d\cos{\theta_d}, v_o \sin{\theta_o}-v_d\sin{\theta_d}]^T\\
&\quad+\gamma_{pv}(\lvert \lvert \Delta p_{pv} \rvert \rvert-d_{pv}) \\
&=\frac{\Delta p_{pv}^{T} }{\lvert \lvert \Delta p_{pv} \rvert \rvert } [v_o \cos{\theta_o}, v_o \sin{\theta_o}]^T-\frac{\Delta p_{pv}^{T} }{\lvert \lvert \Delta p_{pv} \rvert \rvert } [v_d\cos{\theta_d}, v_d\sin{\theta_d}]^T\\
&\quad+\gamma_{pv}(\lvert \lvert \Delta p_{pv} \rvert \rvert-d_{pv}) \\
&=\frac{\Delta p_{pv}^{T} }{\lvert \lvert \Delta p_{pv} \rvert \rvert } u_{pv}-\frac{\Delta p_{pv}^{T} }{\lvert \lvert \Delta p_{pv} \rvert \rvert } [v_d\cos{\theta_d}, v_d\sin{\theta_d}]^T\\
&\quad+\gamma_{pv}(\lvert \lvert \Delta p_{pv} \rvert \rvert-d_{pv}) \\
&=-A_{pv}u_{pv}+b_{pv} \ge 0
\end{split}
\end{equation}
where
\begin{equation}
\begin{cases}
A_{pv}=\frac{\Delta p_{pv}^{T} }{\lvert \lvert \Delta p_{pv} \rvert \rvert }\\
b_{pv}=\frac{\Delta p_{pv}^{T} }{\lvert \lvert \Delta p_{pv} \rvert \rvert } [v_d\cos{\theta_d}, v_d\sin{\theta_d}]^T +\gamma_{pv}(\lvert \lvert \Delta p_{pv} \rvert \rvert-d_{pv})\\
u_{pv} = [v_o \cos{\theta_o}, v_o \sin{\theta_o}]^T
\end{cases}
\end{equation}
According to \textbf{Theorem 1}, if $u_{pv}$  satisfies (\ref{Au2}) and $\lvert\lvert u_{pv}\rvert\rvert \leq u_{pv}^{max}$, then $h_{pv}(x)$ is a valid CBF and the safe set $C_{pv}$ is forward invariant, ensuring the capture-avoidance of the OR. With an appropriate selection of $\gamma_{pv}$, one can make $u_{pv}$ satisfies (\ref{Au2}).

\section{Simulations and Experiments \label{experiment}}
\begin{figure}[h]
     \centering
     \includegraphics[width=8.5 cm]{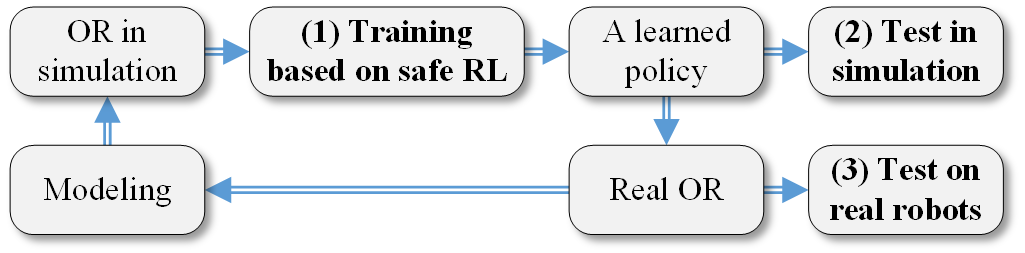}
     \caption{Flow diagram of simulations and experiments}
     \label{experiment_flow}
 \end{figure}

The proposed safe RL framework is evaluated based on a system of robots demonstrated in section \ref{pro_formu} in simulations and experiments. The flow diagram of the simulations and experiments is expressed in Fig. \ref{experiment_flow}. The environment of the simulations and experiments is shown in Fig. \ref{fig:experiment}. The key steps of the simulations and experiments, including (1) training based on safe RL, (2) a test in simulation, and (3) a test on real robots, are presented in detail.

\textbf{Setups.} 
In the simulations and experiment, the DDR pursues the OR following a Nash strategy according to \cite{Ruiz2013Time-OptimalRobot}. The translational velocity of the DDR is $v_d$ = 0.2000 m/s. The translational velocity of the obstacles is $v_{cx}$ = 0.0200 m/s and $v_{cy}$ = 0.0000 m/s. The translational velocity of the OR is $v_o$ = 0.1574 m/s. According to \cite{Ruiz2013Time-OptimalRobot}, the OR can avoid being captured following a Nash strategy, suggesting that it is possible for the OR to achieve objectives covering more than capture-avoidance. The capture distance between the DDR and the OR is $d_{pv}$ = 0.2000 m. The collision distance between the OR and the obstacle is $d_{oc}$ = 0.2000 m.

To train a learned policy for the OR based on the developed safe RL framework, the TD3 algorithm \cite{pmlr-v80-fujimoto18a} is combined with the CBFs of the OR obtained according to (\ref{h_obstacle}) and (\ref{h_pv}). The learned policy is combined with the CBFs in tests in simulation and on real robots then. The coefficients included in CBFs are $\gamma_{oc}$ = 1.0000 and $\gamma_{pv}$ = 1.2000.




\textbf{Training based on safe RL.}
To guide the OR to learn a policy for reaching a randomly generated target position, the reward function of RL is defined as
\begin{equation}
    r= 
    \begin{cases}
        +1000, & \text{$d_1 \le d_{t}$}\\
        -0.01 d_1, & \text{$d_1 > d_{t}$ } \\
    \end{cases}
    \label{rewardset}
\end{equation}
where $d_1$ is the distance between the OR and a randomly generated target position. $d_t$ = 0.0500 m denotes the critical distance of reaching the target position.
In the training, the learning rate of both the critic network and the actor network is $3.0 \times 10^{-4}$. The discount factor is set to 0.99. The number of time steps of an episode is 5000. The number of episodes is 200. The batch size is 256. 

\begin{figure}[h]
     \centering
     \includegraphics[width=7.5cm]{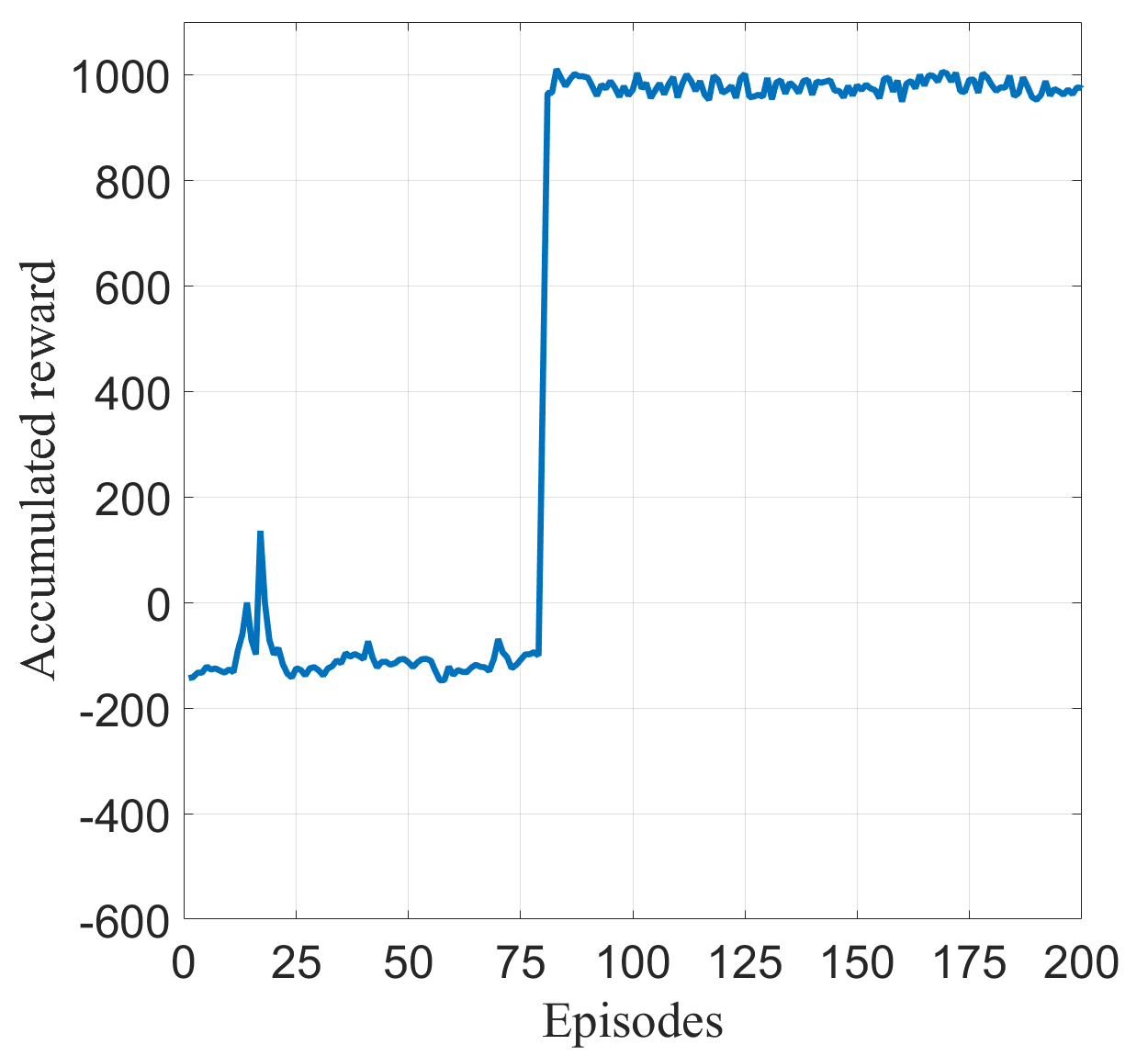}
     \caption{Accumulated reward achieved by the OR in training}
     \label{reward}
\end{figure}

The accumulated rewards achieved by the OR in the training are shown in Fig. \ref{reward}. It is shown that the accumulated rewards converge to 1000 in 80 episodes, suggesting that the OR learned an effective policy for reaching a randomly generated target position. 
If the OR can avoid being captured and avoid the obstacle in an episode, the episode is defined as a safe episode. The number of safe episodes is summarized in Table I. It is shown that all episodes in the training are safe episodes, suggesting that the developed safe RL can ensure the safety of the OR being pursued but with the objective of reaching a target position.

\vspace{-0.2cm}

\begin{table}[h]
\begin{center}
\caption{The number of safe episodes in training}
\begin{tabular}{|c|c|lll}
\cline{1-2}
                   & Value &  &  &  \\ \cline{1-2}
Number of episodes   & 200     &  &  &  \\ \cline{1-2}
Number of safe episodes   & 200     &  &  &  \\ \cline{1-2}
Safety ratio         & 100\%   &  &  &  \\ \cline{1-2}
\end{tabular}
\end{center}
\end{table}
\vspace{-0.2cm}

\textbf{Tests in simulation and on real robots.} 
To evaluate the performance of the combination of the learned policy and CBFs, tests are conducted in simulation and on real robots. The initial positions of the OR and the obstacle are (0.0 m, 0.0 m) and (1.5 m, 0.0 m), respectively. The initial pose of the DDR is (1.0 m, -0.5 m, 0.0 deg). The target position is (2.5 m, 0.0 m).

\begin{figure}[h]
     \centering
     \includegraphics[width=8.5cm]{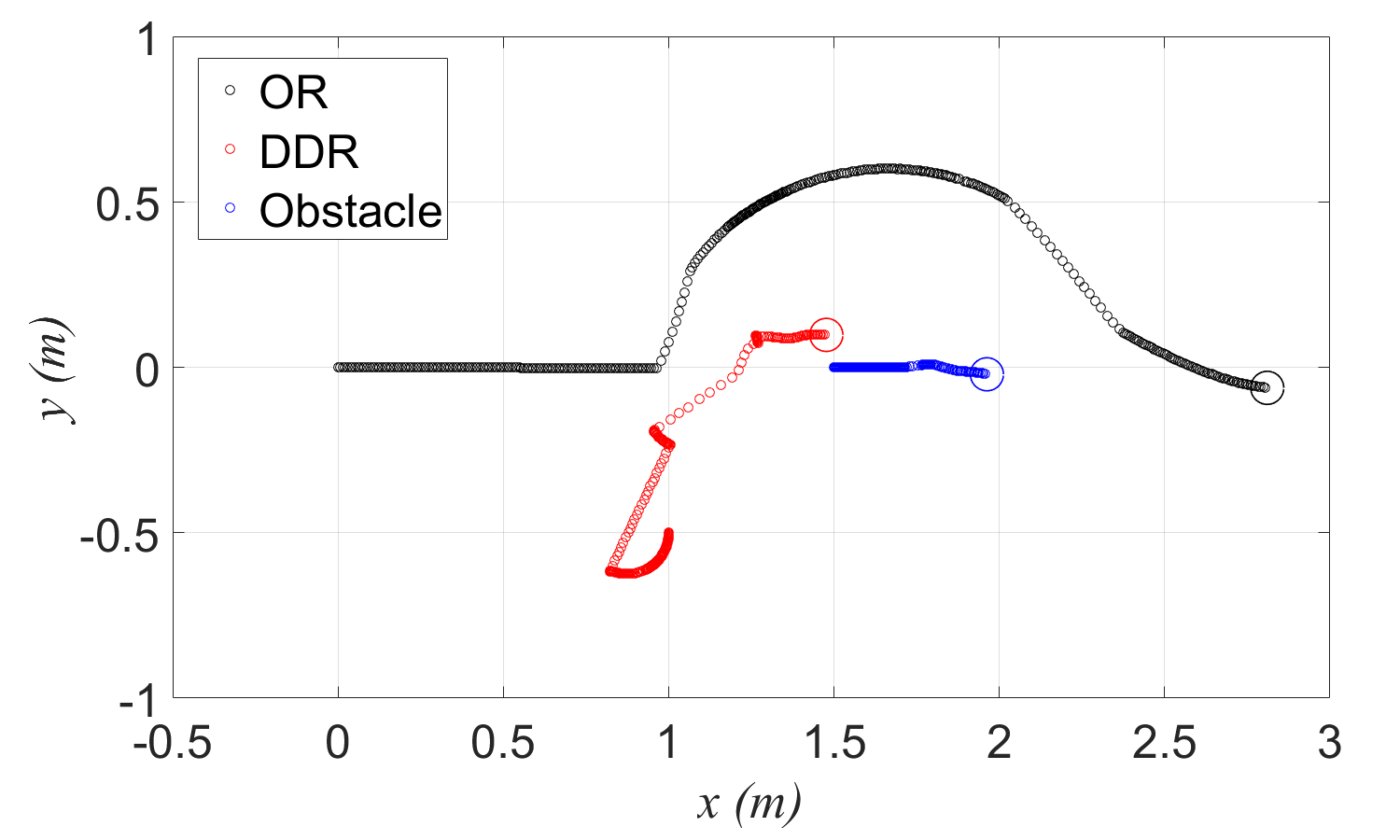}
     \caption{Trajectories of the OR, the pursuing DDR, and the obstacle in a test in simulation}
     \label{simu_trajectries}
\end{figure}

\begin{figure}[h]
     \centering
     \includegraphics[width=8.5cm]{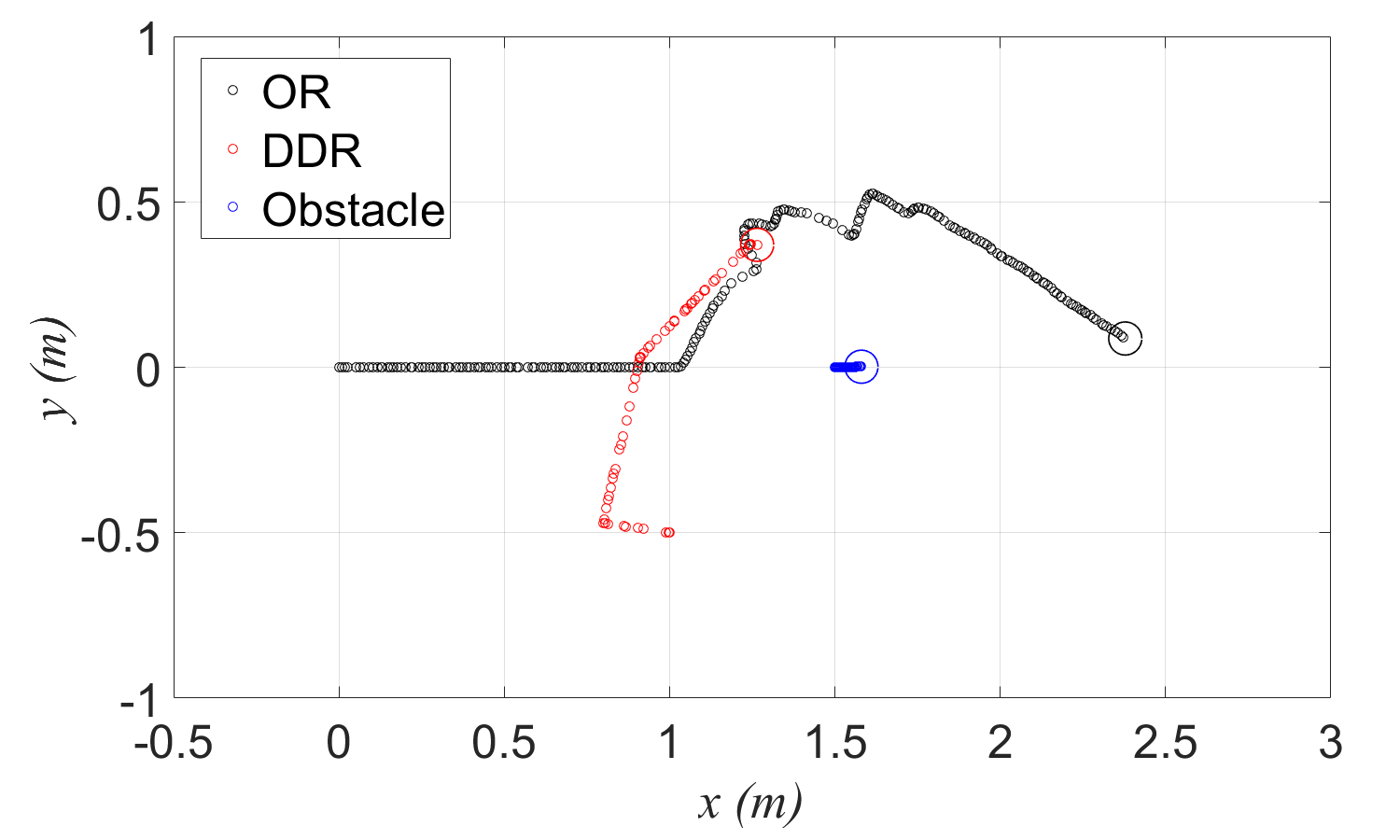}
     \caption{Trajectories of the OR, the pursuing DDR, and the obstacle in a test on real robots}
     \label{real_trajectory}
\end{figure}

The trajectories of the OR, the DDR, and the obstacle in a test in simulation are shown in Fig. \ref{simu_trajectries}. It is shown that the OR can move smoothly to avoid being captured as well as bypass the obstacle and reach the target position finally.
The trajectories of the OR, the DDR, and the obstacle in a test on real robots are shown in Fig. \ref{real_trajectory}. As shown in Fig. \ref{real_trajectory}, the OR can avoid being captured and bypass the obstacle as well. The OR in real-world reaches the target position also. However, 
The trajectories of the robots in real-world are not exactly the same as the trajectories of the robots in simulation. The reason for this difference can be the reality gap \cite{Jakobi1995NoiseRobotics}. The trajectory of the OR shows that the CBF of the OR corrects the action of the OR several times to avoid the obstacle, reflecting the advantage of the CBFs included in the developed safe RL framework in safety. According to the results of tests in simulation and on real robots, it is shown that the combination of the learned policy and CBFs enable the OR to reach a target position and can ensure the safety of the OR both in simulation and on real robots.
\vspace{-0.2cm}
\vspace{-0.2cm}

\section{Conclusions and Future Work \label{summary}}

This paper proposed a safe RL problem for robots inspired by the safety issue of a learned policy for a self-driven vehicle suffering vehicles with unpredictable aggressive behaviors. To address the safe RL problem, a safe RL framework integrating RL with CBFs was developed for robots being pursued but with objectives covering more than capture-avoidance. Based on a system of an OR, a pursuing DDR, and obstacles, the CBFs were designed for the OR.
Simulation and experiments were conducted to evaluate the developed safe RL framework. The results of the simulations and experiments verify the effectiveness of the developed safe RL framework and that of the CBFs of the OR. In the future, we plan to apply the developed safe RL framework to robots such as quadrotors.






\bibliographystyle{IEEEtran}
\bibliography{references}

\end{document}